\documentclass[lettersize,journal]{IEEEtran}

\usepackage{amsmath,amsfonts}
\usepackage{algorithmic}
\usepackage{array}
\usepackage[caption=false,font=normalsize,labelfont=sf,textfont=sf]{subfig}
\usepackage{textcomp}
\usepackage{stfloats}
\usepackage[hyphens]{url}

\usepackage{hyperref}
\usepackage[hyphenbreaks]{breakurl}
\usepackage{verbatim}
\usepackage{graphicx}
\def\BibTeX{{\rm B\kern-.05em{\sc i\kern-.025em b}\kern-.08em
		T\kern-.1667em\lower.7ex\hbox{E}\kern-.125emX}}
\usepackage{balance}

\usepackage{cite}
\usepackage{xcolor}
\definecolor{darkred}{rgb}{.6,0,0}
\definecolor{darkgreen}{rgb}{0,.4,0}
\definecolor{darkblue}{rgb}{0,0,.6}

\usepackage{enumitem}
\usepackage{tikz}
\usepackage{nicematrix}
\NiceMatrixOptions{
	code-for-first-row = \color{darkred} ,
	code-for-last-row = \color{darkred} ,
	code-for-first-col = \color{darkblue} ,
	code-for-last-col = \color{darkblue}
}

\usepackage{orcidlink}

\newcommand{\makeparafit}{\looseness=-1}

\newcommand{\numc}{\ensuremath{\mathsf{n}}}
\newcommand{\numusers}{\ensuremath{\mathsf{m}}}

\newcommand{\user}[1]{\ensuremath{{U}_{#1}}}

\newcommand{\gradvec}{\ensuremath{\vec{\bf G}_{\numusers \times \numc}}}

\newcommand{\ViewMed}{\ensuremath{\vec{\bf V}_{\sf SecMed}}}
\newcommand{\ViewPear}{\ensuremath{\vec{\bf V}_{\sf SecPear}}}
\newcommand{\ViewAgg}{\ensuremath{\vec{\bf V}_{\sf SecAgg}}}

\newcommand{\gvec}[1]{\ensuremath{\vec{\bf g}_{#1}}}
\newcommand{\gvecval}[2]{\ensuremath{{g}_{#1}^{#2}}}

\newcommand{\medv}[1]{\ensuremath{{r}_{#1}}}
\newcommand{\pearv}[1]{\ensuremath{{p}_{#1}}}
\newcommand{\aggv}[1]{\ensuremath{{s}_{#1}}}

\newcommand{\diffmul}[2]{\ensuremath{{\delta}_{#1}^{#2}}}
\newcommand{\diffadd}[2]{\ensuremath{{\Delta}^{#1}_{#2}}}

\begin{document}

\title{Comments on ``Privacy-Enhanced Federated Learning Against Poisoning Adversaries"$^{*}$\thanks{$^{*}$Please cite the journal version of this work published at IEEE TIFS’23~\cite{SchneiderSY23}.}}

\author{
Thomas Schneider\orcidlink{0000-0001-8090-1316}, 
Ajith Suresh\orcidlink{0000-0002-5164-7758}, and 
Hossein Yalame\orcidlink{0000-0001-6438-534X}
\\
Technical University of Darmstadt, Germany
}

\maketitle

\begin{abstract}
In August 2021, Liu~et~al.~(IEEE TIFS'21) proposed a privacy-enhanced framework named PEFL  to efficiently detect poisoning behaviours in Federated Learning (FL) using homomorphic encryption. In this article, we show that PEFL does not preserve privacy. In particular, we illustrate that PEFL reveals the entire gradient vector of all users in clear to one of the participating entities, thereby violating privacy. Furthermore, we clearly show that an immediate fix for this issue is still insufficient to achieve privacy by pointing out multiple flaws in the proposed system.
\end{abstract}


\section{Introduction}
\label{sec:introduction}
Federated Learning (FL) is a new distributed machine learning approach that allows multiple entities to jointly train a model without sharing their private and sensitive local datasets with others. In FL, clients locally train models using their local training data, then send model updates to a central aggregator, which merges them into a global model. FL is used in a variety of applications such as word prediction for mobile keyboards in GBoard~\cite{google_federated_learning} and medical imaging~\cite{medical}. Despite its benefits, FL has been shown to be susceptible to model poisoning~\cite{FLAME} and inference attacks~\cite{SAFELearn}. In model poisoning attacks, an adversary injects poisoned model updates by corrupting a subset of clients, with which the adversary can compromise the user's data privacy as well as the FL model's integrity~\cite{HitajAP17,GeipingBD020}. The recent work of Liu et al.~\cite{LiuLXCHL21} proposed a privacy-enhanced framework called PEFL to detect poisoning behaviors in FL. PEFL aims to prevent malicious users from inferring memberships by uploading malicious gradients and semi-honest servers from invading users' privacy. Furthermore, PEFL claims to be the first effort to detect poisoning behaviors in FL while using ciphertext and uses homomorphic encryption (HE) as the underlying technology.

In this article, we have a closer look at the PEFL system of~\cite{LiuLXCHL21} and identify multiple privacy vulnerabilities. In particular, we show that each of the three main protocols in PEFL -- SecMed, SecPear, and SecAgg, reveals significant information about the user's gradients to one of the computing servers thereby compromising privacy. Furthermore, we demonstrate that combining information from the protocols enables a computation server to learn the gradient vectors of all users in clear, thereby breaking the PEFL system.

\section{Liu et al.’s protocols are not private}
\label{sec:protocols}
In this section, we revisit the Privacy Enhanced Federated Learning~(PEFL) system in~\cite{LiuLXCHL21}, but with our notations for clarity. We begin with an overview of PEFL's four entities:
\begin{itemize}[topsep=3pt,parsep=3pt,partopsep=0pt,leftmargin=10pt,labelwidth=6pt,labelsep=4pt]
    \item \emph{Key Generation Center (KGC):} Trusted entity managing public and private keys $(pk, sk)$ for HE. 
    \item \emph{Data Owners ($\user{x}$):} Each data owner $\user{x}$, for $x \in [\numusers]$, locally trains the local model on their data and computes the gradient vector $\gvec{x} = \{\gvecval{x}{1}, \ldots, \gvecval{x}{\numc}\}$. Here, $\numusers$ denotes the total number of users in the system and $\numc$ denotes the dimension of the gradient vector.
    \item \emph{Service Provider (SP):} SP receives all gradients submitted by data owners and aggregates them (usually by averaging) to produce an optimized global model.
    \item \emph{Cloud Platform (CP):} CP assists SP in the computations and operates on a pay-per-use basis.
\end{itemize}
The threat model assumes that SP and CP are both semi-honest, whereas data owners can be maliciously corrupt. Furthermore, the four entities mentioned above are non-colluding.

We now examine the PEFL system in-depth, focusing on the amount of information visible to each entity. More specifically, we are interested in how much information the cloud platform (CP) learns from the protocol execution.

\subsection{Calculation of Gradients}
The protocol begins with each user $\user{x}$ training the model locally and obtaining the corresponding gradient vector $\gvec{x} = \{\gvecval{x}{1}, \ldots, \gvecval{x}{\numc}\}$. User $\user{x}$ then encrypts and sends the gradient vector to SP using CP's public key $pk_{c}$. As shown in \eqref{eq:grad_view_CP}, the gradient vectors corresponding to all users can be viewed as a matrix $\gradvec$.

\begin{equation}
\gradvec =
    \begin{pNiceMatrix}%
    [margin,
    name=mymatrix,
    first-row,
    first-col,
    nullify-dots,
    xdots/line-style=loosely dotted,]
              &        c_1     & c_2            & \Cdots & c_i            & \Cdots  & c_{\numc-1}           & c_{\numc}          \\[1em]
    \user{1}  & \gvecval{1}{1} & \gvecval{1}{2} & \Cdots & \gvecval{1}{i} & \Cdots  & \gvecval{1}{\numc-1}  & \gvecval{1}{\numc}  \\[0.5em]
    \user{2}  & \gvecval{2}{1} & \gvecval{2}{2} & \Cdots & \gvecval{2}{i} & \Cdots  & \gvecval{2}{\numc-1}  & \gvecval{2}{\numc}  \\[0.5em]
    \Vdots    & \Vdots         & \Ddots         & \Cdots & \Vdots         & \Ddots  &                       & \Vdots     \\
              &                &                &        &                &         &                       &            \\
    \user{\numusers} & \gvecval{\numusers}{1}   & \gvecval{\numusers}{2}   & \Cdots    & \gvecval{\numusers}{i}  & \Cdots  & \gvecval{\numusers}{\numc-1}  & \gvecval{\numusers}{\numc}
    \end{pNiceMatrix}.
    \label{eq:grad_view_CP}
\end{equation}
%

\subsection{Median Computation using SecMed}
The SP and CP use the SecMed algorithm (cf. Figure~4 in~\cite{LiuLXCHL21}) to compute the median value for each of the $\numc$ coordinates. SP sends $\gvecval{j}{i} + \medv{i}$ to CP for each user $\user{j}$ corresponding to a coordinate $c_i$, where $\medv{i}$ denotes a random pad sampled by SP for each coordinate $c_i$ (but the same for all users). The CP decrypts and computes the medians based on these padded values. The CP then encrypts the medians before sending them to the SP. Finally, the SP removes the pad $\medv{i}$ to achieve the desired results $\gvec{y}$ by utilizing the underlying encryption scheme's homomorphic property.

\paragraph*{\bf Leakage}
The view of CP while executing SecMed is consolidated in the matrix $\ViewMed$:
\begin{equation}
\ViewMed =
    \begin{pNiceMatrix}%
    [margin,
    name=mymatrixb,
    first-row,
    first-col,
    nullify-dots,
    xdots/line-style=loosely dotted,]
              &        c_1     & \Cdots & c_i            & \Cdots            & c_{\numc}          \\[0.3em]
    \user{1}  & \gvecval{1}{1} + \medv{1} & \Cdots & \gvecval{1}{i} + \medv{i} & \Cdots    & \gvecval{1}{\numc} + \medv{\numc}  \\[0.5em]
    \Vdots    & \Vdots         & \Cdots & \Vdots         & \Cdots    & \Vdots     \\
    \user{\numusers} & \gvecval{\numusers}{1} + \medv{1}    & \Cdots    & \gvecval{\numusers}{i} + \medv{i}  & \Cdots   & \gvecval{\numusers}{\numc} + \medv{\numc}
    \end{pNiceMatrix}.
    \label{eq:SecMed_view_CP}
\end{equation}
We observe that for each coordinate $c_i$, CP learns a ``shifted" distribution of gradients in clear across all users. This is clearly a violation of privacy~\cite{HitajAP17,GeipingBD020} because it leaks a lot more information to CP and thus does not meet the design goal of `Privacy' claimed in~\cite{LiuLXCHL21}. The main source of the leakage is that SP uses \emph{ same random pad $\medv{i}$ for all users} with respect to a coordinate $c_i$. While the aforementioned leakage could be prevented by using different random pads across users, we emphasize that the use of the same pad is unavoidable for the SecMed algorithm to remain correct. In detail, the median of $\gvecval{j}{i}$ values is calculated by first computing the median of $\gvecval{j}{i} + \medv{i}$, then removing $\medv{i}$ from the result. This requires that the same $\medv{i}$ value be associated with each $\gvecval{j}{i}$ value, or the computation's correctness will be violated.

\subsection{Computing Pearson correlation coefficient using SecPear}
\makeparafit
Once the coordinate-wise medians are computed, the next step in PEFL is to calculate the Pearson correlation coefficient $\rho_{x,y}$ between the coordinate-wise medians $\gvec{y}$ and the gradient of the user $\user{x}$. 
This is achieved via the SecPear protocol (cf. Figure~5 in~\cite{LiuLXCHL21}) where SP communicates $\gvec{x} \cdot \pearv{x}$ and $\gvec{y} \cdot \pearv{y}$ to CP. The view of CP in SecPear with respect to $\gradvec$ is $\ViewPear$:

\begin{equation}
\ViewPear =
    \begin{pNiceMatrix}%
    [margin,
    name=mymatrixc,
    first-row,
    first-col,
    nullify-dots,
    xdots/line-style=loosely dotted,]
              &        c_1     & \Cdots & c_i            & \Cdots            & c_{\numc}          \\[0.3em]
    \user{1}  & \gvecval{1}{1} \cdot \pearv{1} & \Cdots & \gvecval{1}{i} \cdot \pearv{1} & \Cdots    & \gvecval{1}{\numc} \cdot \pearv{1}  \\[0.5em]
    \Vdots    & \Vdots         & \Cdots & \Vdots         & \Cdots    & \Vdots     \\
    \user{\numusers} & \gvecval{\numusers}{1} \cdot \pearv{\numusers}    & \Cdots    & \gvecval{\numusers}{i} \cdot \pearv{\numusers}  & \Cdots   & \gvecval{\numusers}{\numc} \cdot \pearv{\numusers}
    \end{pNiceMatrix}.
    \label{eq:SecPear_view_CP}
\end{equation}
%

\paragraph*{\bf Leakage}
Similar to the problem with SecMed above, here CP learns the correlation between each coordinate in the gradient vector $\gvec{x}$. SP uses the \emph{same random pad $\pearv{x}$ for all coordinates}, which causes leakage. However, using different pads for the coordinates does not address the issue since the use of the same pad is required for the SecPear algorithm to remain correct (cf. Proposition 1 in~\cite{LiuLXCHL21}). More elaborately, for $d_x = \gvec{x} \cdot \pearv{x}$ and $d_y = \gvec{y} \cdot \pearv{y}$, computation of $\rho_{x,y}$ involves computing the covariance $Cov(d_x, d_y)$ and the standard deviations $\sigma(d_x)$ and $\sigma(d_y)$. As shown in Proposition~1 in~\cite{LiuLXCHL21}, the correctness of $\rho_{x,y}$ relies on the following observations:
\begin{gather}
Cov(d_x, d_y) = \pearv{x}\pearv{y} \cdot Cov(\gvec{x}, \gvec{y}),\nonumber \\
\sigma(d_x) = \pearv{x} \cdot \sigma(\gvec{x})~,~
\sigma(d_y) = \pearv{y} \cdot \sigma(\gvec{y}). \nonumber
\end{gather}
If different pads are used for the coordinates, the above relations do not hold, and hence $\rho_{d_x,d_y} = \frac{Cov(d_x,d_y)}{\sigma(d_x)\sigma(d_y)} \ne \rho_{x,y}$.

\subsection{Aggregating the gradients using SecAgg}
SecAgg, the final stage in PEFL, aggregates the gradients after scaling them with a factor based on the Pearson correlation coefficient calculated in SecPear. SP communicates $\gvec{x} + \aggv{x}$ to CP for this purpose, as shown in $\ViewAgg$:

\begin{equation}
\ViewAgg =
    \begin{pNiceMatrix}%
    [margin,
    name=mymatrixd,
    first-row,
    first-col,
    nullify-dots,
    xdots/line-style=loosely dotted,]
              &        c_1     & \Cdots & c_i            & \Cdots            & c_{\numc}          \\[0.3em]
    \user{1}  & \gvecval{1}{1} + \aggv{1} & \Cdots & \gvecval{1}{i} + \aggv{1} & \Cdots    & \gvecval{1}{\numc} + \aggv{1}  \\[0.5em]
    \Vdots    & \Vdots         & \Cdots & \Vdots         & \Cdots    & \Vdots     \\
    \user{\numusers} & \gvecval{\numusers}{1} + \aggv{\numusers}    & \Cdots    & \gvecval{\numusers}{i} + \aggv{\numusers}  & \Cdots   & \gvecval{\numusers}{\numc} + \aggv{\numusers}
    \end{pNiceMatrix}.
    \label{eq:SecAgg_view_CP}
\end{equation}
%

\paragraph*{\bf Leakage}
Again, similar to SecMed, $\ViewAgg$ reveals a ``shifted" distribution of each user's gradient values across all the coordinates to CP. When combining information from $\ViewPear$~\eqref{eq:SecPear_view_CP} and $\ViewAgg$~\eqref{eq:SecAgg_view_CP}, a more significant leakage occurs. 
Consider the gradient at coordinates $i,j$ for user $\user{x}$. From $\ViewPear$, CP learns $a_1 = \gvecval{x}{i} \cdot \pearv{x}$ and $a_2 = \gvecval{x}{j} \cdot \pearv{x} = (\gvecval{x}{i} \cdot \diffmul{x}{ij}) \cdot \pearv{x}$ where $\diffmul{x}{ij} = \gvecval{x}{j}/\gvecval{x}{i}$. Similarly, CP learns $b_1 = \gvecval{x}{i} + \aggv{x}$ and $b_2 = \gvecval{x}{j} + \aggv{x} = (\gvecval{x}{i} + \diffadd{x}{ij}) + \aggv{x}$ from $\ViewAgg$, where $\diffadd{x}{ij} = \gvecval{x}{j} - \gvecval{x}{i}$. Given that CP can compute both $\diffmul{x}{ij}$ and $\diffadd{x}{ij}$ in clear, CP learns $\gvecval{x}{i}$ and $\gvecval{x}{j}$ by solving the equations. For instance, $a_2 = (\gvecval{x}{i} + \diffadd{x}{ij}) \cdot \pearv{x} = a_1 + \diffadd{x}{ij} \cdot \pearv{x}$ reveals $\pearv{x}$. Using this method, CP learns the entire gradient matrix $\gradvec$, thereby breaching the PEFL system's privacy. 

\subsection{Practical and probabilistic attacks}
Another practical attack on PEFL would be to allow CP to register as an honest user $\user{\numusers+1}$ in the PEFL system and submit its gradients. This action does not violate the semi-honest assumption of CP in the PEFL threat model and may represent scenarios in which CP has some side channel information about some user gradients. Knowing $\gvec{\numusers + 1}$, CP learns $\medv{i}$ for all $i \in [\numc]$ and hence the gradient matrix $\gradvec$ given in \eqref{eq:grad_view_CP} in clear from \eqref{eq:SecMed_view_CP} and thereby breaking the privacy of the entire PEFL system.

Considering the matrices $\ViewMed$ and $\ViewPear$ together, we note that it is sufficient for CP to be aware of just one gradient, say $\gvecval{x}{i}$, to compromise the system's privacy. In particular, $\gvecval{x}{i}$ will allow CP to learn the random pad $\medv{i}$ corresponding to the $i$-th coordinate $c_i$ in $\ViewMed$, revealing the gradients of all users at that coordinate. 
Similarly, knowing $\gvecval{x}{i}$ allows CP to learn the random pad $\pearv{x}$ corresponding to user $\user{x}$ in $\ViewPear$ and reveals the gradient vector $\gvec{x}$ to CP in clear. CP will now learn the entire gradient matrix $\gradvec$ by combining the information from $\ViewMed$ and $\ViewPear$.

Finally, we note that CP can launch probabilistic attacks by looking for similar values in the matrices $\ViewMed$ and $\ViewPear$ and attempting to correlate the random pads. This is possible in PEFL because CP is aware of the correlation between different rows of $\ViewMed$ as well as columns of $\ViewPear$.

\section*{Additional Remarks}
Although our privacy issues mentioned in Section~\ref{sec:protocols} have been published in January 2023~\cite{SchneiderSY23}, several subsequent papers continued to reference~\cite{LiuLXCHL21} as a potential solution for private federated learning.\footnote{As of September 25, 2024,~\cite{LiuLXCHL21} has been cited $206$ times according to Google Scholar.} 
While a few works, e.g.~\cite{SP:GehlharM0SWY23,CS:XuL23,TIFS:WuZL24}, have acknowledged the privacy concerns we raised, several of subsequent works either propagate these errors or adopt the constructions from~\cite{LiuLXCHL21}, thereby unintentionally inheriting the same privacy vulnerabilities~\cite{ACCESS:XiaCYM23,CSUR:CYLZXU24,COMSUR:WanQNXGH24,TSC:ZZWZLL24}.
We believe this oversight is partly due to the limited visibility of our comments paper at TIFS'23~\cite{SchneiderSY23}. Consequently, to prevent the continued propagation of the flawed algorithms in~\cite{LiuLXCHL21} into future research, we also put this article to an ePrint.

\section*{ACKNOWLEDGEMENTS}
This project received funding from the European Research Council~(ERC) under the European Union's Horizon 2020 research and innovation program~(grant agreement No.850990 PSOTI) and was co-funded by the Deutsche Forschungsgemeinschaft (DFG) -- SFB1119 CROSSING/236615297 and GRK2050 Privacy \& Trust/251805230.

\bibliographystyle{IEEEtranS}
\bibliography{References}

\end{document}